\documentclass{article}
\usepackage{ijcai22}

\usepackage{times}
\usepackage{soul}
\usepackage{url}
\usepackage[hidelinks]{hyperref}
\usepackage[utf8]{inputenc}
\usepackage[small]{caption}
\usepackage{graphicx}
\usepackage{amsmath}
\usepackage{amsthm}
\usepackage{bm}
\usepackage{dsfont}
\usepackage{amssymb}
\usepackage{multirow}
\usepackage{fancyhdr}
\usepackage{booktabs}
\usepackage{algorithm}
\usepackage{algorithmic}
\newtheorem{definition}{Definition}
\newtheorem{assumption}{Assumption}

\usepackage{color}

 \def \cO{\mathcal{O}}
 \def \cU{\mathcal{U}}
 \def \cD{\mathcal{D}}
 \def \cI{\mathcal{I}}
 \def \cT{\mathcal{T}}

 \def \P{\mathbb{P}}
\def \bE{\mathbb{E}}
\title{On the Opportunity of Causal Learning in Recommendation Systems: \\Foundation, Estimation, Prediction and Challenges}
\author{
Peng Wu$^{1\ast}$ \and
Haoxuan Li$^1$\footnote{Contributed equally}\and
Yuhao Deng$^{1}$\and
Wenjie Hu$^1$\and
Quanyu Dai$^{2}$\and
Zhenhua Dong$^2$\footnote{Contact Author} \and
Jie Sun$^{3}$\and
Rui Zhang$^{3}$\And 
Xiao-Hua Zhou$^{1\dagger}$\\
\affiliations
$^1$Peking University \\
$^2$Huawei Noah's Ark Lab\\
$^3$Huawei Hong Kong Research Center\\
\emails
azhou@math.pku.edu.cn, 
dongzhenhua@huawei.com
}

\begin{document}
\maketitle

\thispagestyle{fancy}
\lhead{To Appear in IJCAI-ECAI 2022}

\begin{abstract}
Recently, recommender system (RS) based on causal inference has gained much attention in the industrial community, as well as the states of the art performance in many prediction and debiasing tasks. Nevertheless, a unified causal analysis framework has not been established yet. Many causal-based prediction and debiasing studies rarely discuss the causal interpretation of various biases and the rationality of the corresponding causal assumptions. In this paper, we first provide a formal causal analysis framework to survey and unify the existing causal-inspired recommendation methods, which can accommodate different scenarios in RS. Then we propose a new taxonomy and give formal causal definitions of various biases in RS from the perspective of violating the assumptions adopted in causal analysis. Finally, we formalize many debiasing and prediction tasks in RS, and summarize the statistical and machine learning-based causal estimation methods, expecting to provide new research opportunities and perspectives to the causal RS community.  % standard; % based on statistics and machine learning for causal estimand   that can be used in RS scenarios
\end{abstract}

% =======================================================================
\section{Introduction}

% \subsection{Background}
% 1. 因果推断
% Exploring causality is one of the most valuable goals pursued by many disciplines  
%such as artificial intelligence, econometrics, philosophy, cognitive science, epidemiology, public health, and recommender system (RS). 
     
%  The fundamental challenge of causal inference is the mismatch between the inferential power of the collected data and the targeted counterfactual quantity ~\cite{Correa-etal2019}, as data only reflect the association among variables and it is not enough to answer a causal problem without imposing further assumptions. 

% 2. 因果推断与推荐系统的关系：% ,SDM-SaitoSN19 
% 3. 基于因果的推荐方法的优势
% As one of the most concentrated industrial application fields using cutting-edge technology,   
In recent years,  causal inference has attracted extensive attention from both academic and industrial communities. New theories, methods, and applications of causal inference are emerging at an alarming rate. Recommender system (RS) is a promising field for the development and application of causal inference.  Many practical problems of interest in RS are essentially causal problems, such as post-view click-through rate prediction~\cite{Guo-etal2021}, post-click conversion rate prediction~\cite{Zhang-etal2020,Guo-etal2021}, and uplift modeling~\cite{Sato-Singh2019,Sato-Takemori2020}. 
Causal recommendation approaches have several advantages over the traditional recommendation methods, including better interpretability and stability, higher accuracy, and generalization ability.  Its effectiveness has been verified on both numeric experiments and theoretical analyses across strands of literature~\cite{Wang-Zhang-Sun-Qi2019}.
  % \col{The fundamental problem of causal inference is that only one counterfactual is observable~\cite{Rubin1974}.} 

%1.因果框架的重要性
%2.强调讨论因果假设的重要性和我们针对每个假设都给出了对应的情景。。。
%3.针对违背不同的因果假设，从‘因果角度’给出了推荐领域bias的解释
% 4. 当前因果推荐方法的问题 
% A causal analysis framework includes clear causal formulation on the scientific questions of interest, formal definitions of biases, debiasing approaches and associated assumptions. However, the existing causal approaches in RS lack these discussions, inducing the prevalence of many nebulous causal concepts in this field and impeding the development of causal recommendation methods. Much confusion needs to be clarified:  what exactly is being estimated, for what purpose, in which scenario, by which technique, and under what plausible assumptions.  
% In addition, the conditions required for identifiability (or recoverability)  and the assumptions underlying the working models are usually overlooked. 
Nevertheless, a unified causal analysis framework has not been established yet. On one hand,  the existing causal approaches in RS lack a clear causal and mathematical formulation on the scientific questions of interest, inducing the prevalence of many nebulous causal concepts in this field and impeding the development of causal recommendation methods.  
Much confusion needs to be clarified:  what exactly is being estimated, for what purpose, in which scenario, by which technique, and under what plausible assumptions. 
On the other hand, a conspicuous feature of the observational data in RS is the existence of various biases, which is the main obstacle to drawing causal conclusions. However, formal causal definitions of the biases in RS are still not clear, even though miscellaneous biases have been discovered and proposed in a descriptive way~\cite{Chen-etal2020}.    
Due to the lack of formal articulation of causal questions and
types of biases, it is difficult to clearly discuss the theoretical properties, merits, and drawbacks of the debiasing approaches. And it is hard to clearly explain the assumptions underlying the methods. As a consequence, it is hard to develop new debiasing algorithms.

% This shortage further greatly restricts the development of debiasing methods.  
%1.因果框架的重要性
%2.强调讨论因果假设的重要性和我们针对每个假设都给出了对应的情景。。。
%3.针对违背不同的因果假设，从‘因果角度’给出了推荐领域bias的解释
%4.应用该框架/指南于‘许多’RS任务，并且转化为一个因果问题，包含ITE估计，依从性，policy evaluation and learning
%5.综述了一些统计/机器学习/深度学习方法，来估计上述因果量in RS

In this article, we aim to overcome the above limitations by proposing a causal analysis framework for RS within the potential outcome  framework~\cite{Rubin1974,Imbens-Rubin2015}, through which we survey and unify the existing causal-inspired recommendation methods. We provide a causal perspective on biases in RS by analyzing the causal assumptions that are potentially used but not discussed in existing studies, and then discuss the corresponding recommendation scenarios violating the above assumptions. In fact, a large number of recommendation % RS 
tasks are rigorously clarified by applying the proposed causal framework. In addition, we overview statistical and machine learning-based causal estimation methods, providing many opportunities for innovative causal RS research. 
The main contributions of this paper are summarized as follows: 
    \begin{itemize}
        \item Providing a guideline of how to define, recover and estimate a causal estimand in RS, thereby explicating the perplexing causal concepts within the potential outcome framework.   
        
        \item Providing a new taxonomy and giving formal causal definitions of various biases in RS from the perspective of violating what assumptions are adopted in standard causal analysis.
        % and solve a causal problem.  
        
        \item Revealing the key assumptions underlying various debiasing approaches, as well as distinguishing the selection bias and the confounding bias in RS.
        
        \item Applying the proposed causal analysis framework to the classical debiasing and prediction tasks in RS, and summarizing the  statistical and machine learning-based causal estimation methods. 
        
        \item Sharing and discussing several noteworthy open research directions for the causal RS community.
        
        % and giving a principled operational guideline for them.  
        % unifying them into a few causal frameworks.
        % 提供了一个全局的视角，看待因果问题;
    \end{itemize}

% Sato-Singh2019 % Sato-Singh2019

% In this paper, we aim to give a causal analysis framework to accommodate different scenarios in RS, thereby providing a principled and rigorous operational guideline for the causal recommendation. 
  % However, the taxonomy and definition of biases in causal inference are significantly different from those in RS. For example, selection bias and confounding bias are the most common biases in causal inference; the widespread biases in RS contain selection bias, exposure bias, position bias, popularity bias, and conformity bias~\cite{Chen-etal2020}. Besides, the definition of selection bias differs in causal inference and RS. 
 %  The former only have a  descriptive definition and 

% ==========================================================
\section{Causal Analysis Framework in RS} 
\begin{figure}[t]  
 \centering
 \includegraphics[scale = 0.58]{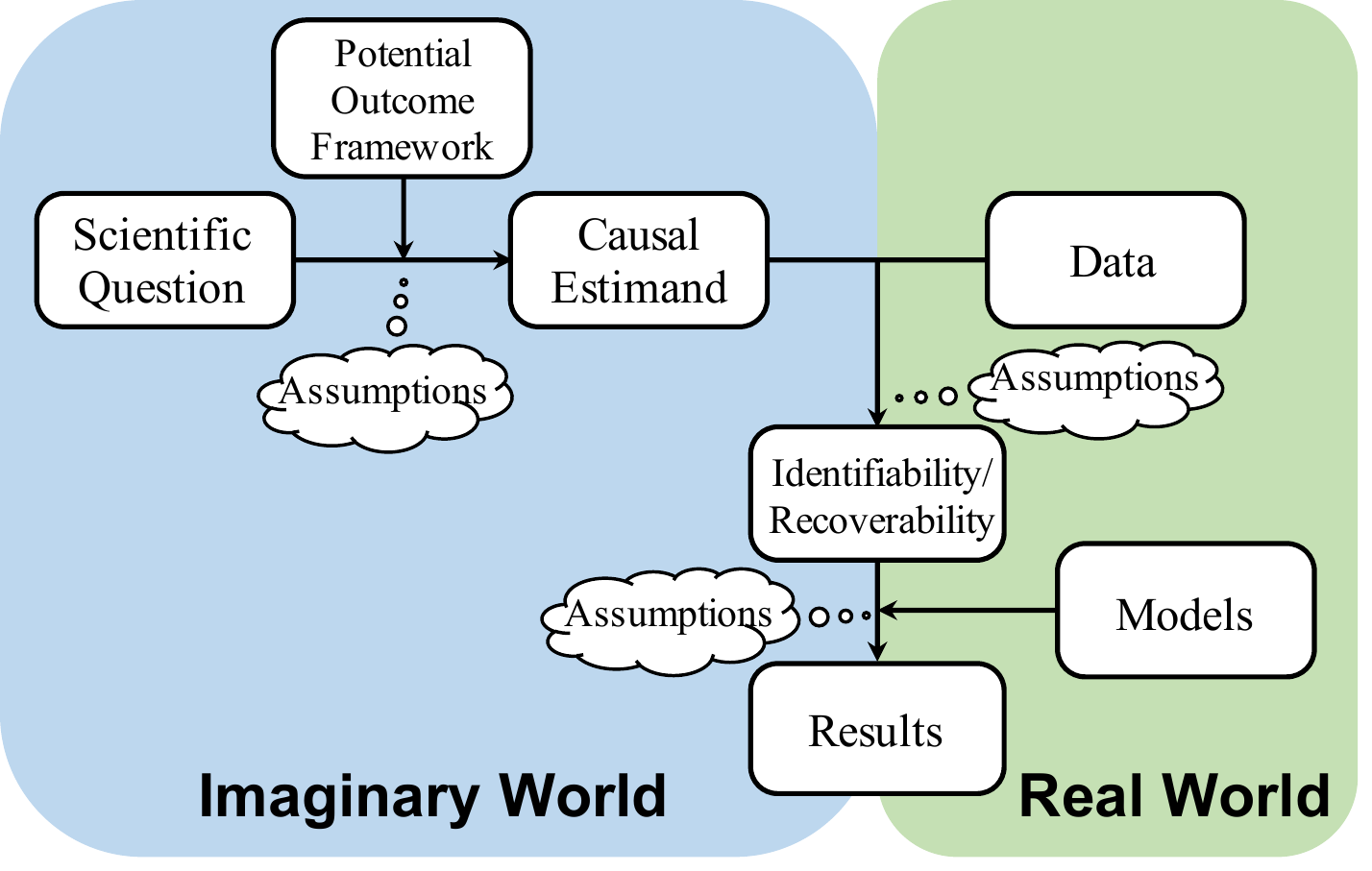}
 \caption{Causal analysis framework in RS.}
  \label{fig1}
\end{figure}

The proposed causal analysis framework refers to a unified workflow of investigating causal problems in RS, which consists of three steps: (1)   Define a causal estimand to answer the scientific question; % under study. 
	(2)  Discuss the recoverability of the estimand given the data; 
	(3) Build models to obtain the consistent estimator of the estimand.  
	Figure \ref{fig1} depicts the causal analysis framework. 

One of the main contributions that causal inference has brought is a focus on clearly defined estimands before building models~\cite{Daniel-etal2016,Vansteelandt-Dukes2021}. Through formalizing the scientific question into a well-defined causal estimand in an imaginary world, we can answer the following questions: what exactly is being estimated and for what purpose.  
Yet, little literature in RS has an explicit statement of the estimand of interest.
  
%This attention has led to a real shift in the statistical approach~\cite{Daniel-etal2016}. 
% The traditional approach of starting a statistical analysis with building a model is increasingly being questioned and abandoned~\cite{Vansteelandt-Dukes2021}.  
% A typical causal inference method is to first translate the scientific question under study into a well-defined causal estimand before adopting a model to estimate it.
  
% Through formalizing the scientific question into a causal estimand, we can answer the following questions: what exactly is being estimated and for what purpose.\col{In Section 3}, 
After defining estimands, we proceed to consider whether a consistent/unbiased estimator, under suitable assumptions, can be derived from the observed data. 
  It is equivalent to theoretically discussing 
   the recoverability property of the estimand~\cite{Mohan-Pearl2021}. 
  If the available RS data does not support deriving a consistent estimator under plausible assumptions, then we should consider how to collect new data; If the estimand is recoverable, we can explicitly present and assess the recoverability assumptions in different RS scenarios underlying the estimation approaches and then build models to estimate it.

As shown in Figure 1, we need a variety of assumptions to climb from association (data and model) to causation (causal estimand and causal conclusion) at each stage of the causal analysis framework. Violating these assumptions may result in various biases. This perspective provides a unified way to discuss the different biases in RS. % and enables us to define them formally.  
Table \ref{tab1} clearly presents the most common assumptions in causal inference and establishes their connections to the biases in RS, from which we can define the descriptive biases in RS formally using the rigorous syntax of causal inference. In addition, it also provides an opportunity to apply the existing causal inference methods to RS. For example, the non-compliance problem and interference bias have been intensively studied in causal inference literature, while rarely being discussed in RS. 
% In Section 4, we will define the biases in RS formally from a new perspective of violating what assumptions adopted in the causal analysis framework.    
% Figure 2 present various biases in RS and in causal inference. Through establishing the connections among them, 
	\begin{table*} 
 %\footnotesize
 \vspace{-0.3cm}
 \small
\centering
\begin{tabular}{llll}    
\toprule
    & Assumptions   &   Biases in causal inference   &    Biases in   ~\cite{Chen-etal2020}  \\
\midrule	
 \multirow{2}{*}{Define causal estimands}   &  SUTVA(a)        &   undefined   & position bias   \\
                        &  SUTVA(b)    & interference bias    & conformity bias  \\ \midrule
\multirow{5}{*}{Recoverability}                          % &  consistency   &            \\
      & consistency   &  noncompliance   & undefined \\
      & positivity   &  undefined   & exposure bias \\
            & exchangeability   & confounding bias  & popularity bias        \\
              & conditional exchangeability   & hidden confounding bias   & undefined \\     
              & random sampling   & selection bias  &  user/model selection bias, exposure bias         \\
	         \midrule 
	       Model       & model specification   & model mis-specification  & inductive bias            \\   
                  \bottomrule         
	\end{tabular}
	\caption{New perspective of biases in RS.}  \label{tab1}
\end{table*}

 % Therefore, it is %desirable   
%	necessary to compare and relate these biases, for developing new algorithms and  
 % deepening the understanding of the various existing debiasing methods of RS. 

% 6. 两种策略:  一个有偏的数据；一个有偏的数据 + 少量无偏的数据; 
% \col{There are two data collection strategies to deal with various biases in RS.  One is to use a biased (observational or non-uniform) dataset~\cite{Schnabel-Swaminathan2016} and the other is to combine a biased dataset with an unbiased (experimental or uniform) dataset~\cite{Wang-etal2021, Chen-etal2021}.  
% Most debiasing methods in RS are based on a biased dataset. However, a significant barrier of this strategy is the unavoidable problem of hidden confounders, i.e., unmeasured features that affect both treatment and outcome. Hidden confounding will distort the causal estimators and lead to biased conclusions, even if the sample size is infinite~\cite{Kallus-Puli2018}. With a biased dataset, we can never test whether there is hidden confounding.  Nevertheless, a possible remedy can be found if there exists an unbiased dataset. This is because the randomly controlled unbiased data has no hidden confounding, and it provides the gold standard for evaluating the debiasing approaches. To the best of our knowledge, there are no debiasing methods in RS that are designed to deal with the hidden confounding bias. We will discuss it in Section 6. }     

% 7. 不同的数据，不同的科学问题，和不同的偏差类型，对应着不同的因果框架。 
%From a global perspective, 
We emphasize that different causal problems in RS correspond to different causal estimands, and may suffer from different types of biases given the collected data. Then, various assumptions and models are needed to estimate the estimands and answer the scientific problems.   
\section{From Scientific Question to Causal Estimand} 
%\subsection{From association to causation} % : a distributional perspective 
 % nevertheless not limited to this framework. 
In this section, we provide a guideline of how to formalize a causal problem by using the potential outcome framework. 
% The concepts discussed in this paper are very general and also applicable to the structural causal model framework~\cite{Pearl-2009}. % Pearl-Glymour-Jewell2016 
% Throughout, we focus on the most common scenario of fixed treatments which do not vary over time~\cite{Hernan-Robins2020}.  
% Rubin's Causal Model~\cite{Rubin1974} --- also known as the potential outcome framework~\cite{Imbens-Rubin2015}.
 %\pw{but not limited to this framework}.  
% =========================================== 
% \subsection{Key elements of defining a causal estimand}
% \subsection{Key elements of potential outcome framework} 
%Before presenting the basic ingredients of describing a causal problem, 
The workflow of translating a scientific problem into a meaningful causal estimand is summarized as follows: 
	(1) Define the unit; 
	(2) Define the treatment, feature, outcome, and potential outcomes corresponding to the scientific question under study; 
	(3) Define the target population; 
	(4) Define the causal estimand. 
	% to answer the scientific question.  

% particularly in the field of  RS.
The unit is the most fine-grained research subject. Unit is a terminology that is often overlooked in RS. However, a clear explanation of it is very important to define the causal estimand.    
 In RS, a unit usually corresponds to a user-item pair~\cite{Guo-etal2021}; sometimes it is a user~\cite{Liang-Charlin-Blei2020} or an item~\cite{Deldjoo-etal2021}. The variety and vagueness of the unit stem from the fact that RS involves two entangled populations: users and items. Therefore, an explicit statement of the unit is helpful to eliminate ambiguity in interpretation.
For each unit, we have a treatment $T$, an outcome $Y$, and possibly a feature vector $X$.  Usually, $X$ is the attributes or feature embedding of the unit.
However, $T$, $X$, and $Y$ are insufficient to define a causal estimand. The potential outcome is a general syntax to formalize causal estimand, thereby translating the meaningful causal problems into causal parameters~\cite{Goetghebeur-etal2020}. 

\begin{definition}
[Potential outcome] 
A potential (or counterfactual) outcome $Y(t)$ for $t\in \cT$ is the outcome that would be observed if $T$ had been set to $t$.  	
\end{definition}

 % The most fine-grained causal effect, i.e., individualized causal effect, is defined at the unit level. For example, consider the case of binary treatment, namely $\cT = \{0,1\}$,  $Y_{m}(1)$ and $Y_{m}(0)$ denote the potential outcomes for unit $m$, then individualized causal effect is defined as $Y_{m}(1) - Y_{m}(0)$. Unfortunately, one can never calculate the individualized causal effects~\cite{Holland1986,Morgan-Winship-2015},  as only one of $Y_{m}(1)$ and $Y_{m}(0)$ is observable. In practice, the individualized causal effect often refers to the feature-specific causal effect~\cite{Guo-Zhou-Ma2021}, defined by 
 %	$\bE( Y(1) - Y(0)  \mid X = x )$, 
 %	 where $\bE$ means taking an expectation    with respect to the target population. 
  %a population of interest.   
 	 % Because it is typically impossible to calculate individual-level causal effects, we usually attention on the estimation of carefully defined aggregate causal effects. we typically define aggregate causal effects as a average of these individual-level effects. These average causal effects can be defined for nay subset of population. 
 % Therefore, the finest-grained causal effect that can be estimated is the average causal effects conditional on all the observed features $X$, such as  
%	and  
%	 \begin{equation}  \label{eq2}     \bE( Y(1)  \mid X = x ),   \end{equation}  
 % which are two common estimands in RS. 

% SUTVA  
% Rubin (1980)\cite{Rubin1980} proposed an assumption of causal effect stability known as the stable unit treatment value assumption (SUTVA). 
{The stable unit treatment value assumption (SUTVA,~\cite{Rubin1980}) is necessary to ensure the well-definedness of potential outcome $Y(t)$.}  
\begin{assumption}[SUTVA] 
         (a) No multiple versions of treatment,  
      only a single version of the treatment and
a single version of the control; 
     (b) Non-interference, the potential outcomes of a unit are not affected by the treatment status of the other units. 
\end{assumption}

In RS, even though there is already a certain amount of studies based on a causal perspective, most of them tacitly assume that SUTVA holds without discussion. However, in many scenarios, the SUTVA assumption does not necessarily hold. \textcolor{black}{For example, the position bias can be seen as a violation of SUTVA(a).}
{In the task of click-through rate prediction, suppose a unit is a user-item pair. Define $Y_{u,i}(1)$ as the  click behavior if the item $i$ is exposed to the user $u$. Then $Y_{u,i}(1)$ will rely on the position of exposure and multiple versions of treatment occur}.  
 In addition, the conformity bias means that users tend to rate similarly with others in a group. It violates SUTVA(b), i.e., non-interference.  
This is because the conformity phenomenon may lead to the value of $Y_{m}(t)$ depending on the treatment value $T_{j}$ for unit $j \neq m$.  
Violations of SUTVA(b) in RS should receive more attention due to the existence of users' social networks. 
% the violation of SUTVA(b) should be paid more attention to in the RS with the existence of a user social network.
  %For position bias, we can consider the position as a treatment, then it can be addressed as a problem of multi-valued treatment, corresponding to scenario in Section 5.3. 

% When SUTVA (a) is violated, the definition of potential outcome is implicit; examples are provided in Section 4.2. 
% If SUTVA (b) is violated, then the value of $Y_m(t)$ depends on the treatment value $T_j$ for unit $j \neq m$. In such a case, it will suffer from interference problems. % if we disregard it.  
% In this paper, we maintain the SUTVA assumption. 
%In practice, we assume SUTVA(b) to simplify problem of interest. 
 %   \hwj{This article assume SUTVA holds} 
% Regrettably,  
To clarify the population of interest, we need to specify a target population, which is the population that we want to make an inference on.
% \begin{definition} 
% [Target population] 
% Target population is the population that we want to make an inference on. %  the causal conclusion to.   
% \end{definition}is the population that we want to make an inference on.
We denote $\P$ and $\bE$ as the distribution and expectation on the target population.  
In RS, the target population is usually the population consisting of all user-item pairs, or all users, or all items.
% which depends on the specific scientific question. % and the analysis methods.
Based on the target population, we can define the causal estimand as follows.
% represented by the test set.   

\begin{definition}[Causal estimand] 
Causal estimand is a functional of 
% an quantity that is defined in terms of 
the joint distribution of treatment, feature and potential outcomes on the target  population, providing a recipe for answering the scientific question of interest from any hypothetical data whenever it is available~\cite{Pearl2019}.  	
\end{definition}

% The causal estimand involves a distribution of observed variables and potential outcomes, 
  
% elucidate  
% Finally, we emphasize 
It should be noted that the definition of the causal estimand does not involve the data collected %, the distribution of the variables, 
and the model adopted. More detailed examples are provided in Section 5.
% \subsection{Illustrations} % examples  

% ==================================================================
\section{Recoverability: From Causal Estimand to Consistent Estimation}
% 解释识别性
%	 However, the causal estimand concerns the potential outcomes that 
%may be unobservable.  
In this section, we discuss the recoverability of the estimand, and relate the associated assumptions with the biases in RS.  

\begin{definition}
[Recoverability of target quantity $Q$~\cite{Mohan-Pearl2021}] Let $\mathcal{A}$ denote the set of assumptions about the data generation process and let $Q$ be any functional of the underlying distribution $\P(X, T, \{Y(t), t\in \cT\})$. $Q$ is recoverable if there exists a procedure that computes a  consistent estimator of $Q$ for all strictly positive observed-data distributions. 
\end{definition}

% Not all causal estimand can be identified, such as consistent estimate of individualized causal effect is often infeasible. 
Recoverability is a crucial ingredient in causal inference, while it is rarely discussed in RS. The significance of discussing recoverability is at least twofold: First, we can ascertain whether a consistent estimator of the counterfactual estimand can be obtained from the data available under some reasonable assumptions.    
Second, if the estimand is recoverable, we can explicitly present the recoverability assumptions underlying the estimation approaches. 
This provides a desirable perspective to evaluate the debiasing methods by assessing the assumptions and provides an opportunity to develop new approaches by weakening the assumptions.

% 常用假定: 相合性 
% For ease of presentation, we focus on the case of binary treatment.   
% are the prerequisite for identifying the causal estimand. 
% maintained 
\subsection{Common Assumptions for Recoverability}
Consistency and positivity are two indispensable assumptions required in most causal inference approaches for recovering the causal estimand.   
\begin{assumption}[Consistency]  $Y(t) =  \sum_{t^*\in \cT} I(t^*=t) Y$ for any $t \in\cT$.
\end{assumption}

\begin{assumption}[Positivity]  $\P(T=t \mid X = x ) > 0$ for any $t$ and $x$.
\end{assumption}

%   thus possibly enabling us to obtain unbiasedly estimate with observed data.

The consistency assumption implies that $Y_{m}(t) =  Y_{m}$ if $T_{m} = t$ for each unit $m$. It links the potential outcomes in the hypothetical world to the observed outcomes in reality.  Positivity ensures that units have a positive probability to take each treatment, and this assumption is sometimes also called ''overlap'' to depict the features of units overlapping in different treatment groups~\cite{Hernan-Robins2020}.

In RS, exposure bias results from that a user is only exposed to a part of specific items. That is, some users are exposed to some items with zero probability. Therefore, exposure bias can be viewed as a violation of the positivity assumption. Noncompliance problem is prevalent in RS, while rarely discussed. A typical example is the exposure-click-conversion model, where we assume that the exposure affects conversion only through click. If the effect of exposure on conversion is of interest, 
the inconsistency between exposure and click, called noncompliance, would violate the consistency assumption and pose a big challenge in estimating the causal effect.
More discussions are provided in Section 5.5. 
%Thus, the consistency assumption may be violated in setting of noncompliance.}

Confounding bias is often caused by violation of the conditional exchangeability assumption defined as follows. 
\begin{assumption}[Conditional exchangeability] $ Y(t) \perp T \mid X$, for any $t \in \cT$. A stronger version is exchangeability: $ Y(t) \perp T$, for any $t \in \cT$.
\end{assumption}

Conditional exchangeability is also called ignorability or unconfoundedness~\cite{Rosenbaum-Rubin1983}.  
In the language of the causal graphical model, 
conditional exchangeability  means that $X$ blocks every back-door path between $T$ and  
$Y$~\cite{Pearl-Mackenzie2018,Hunermund-Bareinboim2019}.
 Besides, an underlying  assumption in causal inference is 
that the observed samples can reflect the target population. 
% another assumption is required.  
% \pw{To simplify notations, we use $\P$ to denote $\P_{\cO}$ whenever Assumption 5 holds.}  
\begin{assumption}[Random sampling]
$\P(x,t,y) = \P_{\cO}(x,t,y)$, where $\P$ represents the target population distribution and $\P_{\cO}$ represents the observed sample distribution.
\end{assumption}

The combination of Assumptions 1-5 can obtain the recoverability of most causal quantities by using observed samples. For example, if $\bE[ Y(t) \mid X = x ]$ is of interest, we can reformulate it as  
	\begin{equation}  \label{eq3}
\resizebox{.91\linewidth}{!}{$
    \displaystyle
 	\bE[ Y(t) | X = x ]  = \bE[ Y(t) | X = x, T  = t]   = \bE[ Y | X = x, T  = t], 
$}
\end{equation}
where the first identity relies on the positivity and conditional exchangeability assumptions, and the second identity requires the consistency assumption.  Based on (\ref{eq3}), under random sampling assumption, $\bE[ Y| X = x, T  = t]$ can be estimated consistently from the observed data, which has been implemented with satisfactory performance by a large number of RS literature, then the recoverability is realized. 

% Ideal randomized experiments (uniform or unbiased data) can be used to recover the causal effects because the randomized assignment of treatment leads to exchangeability. As a consequence, the observed association can reflect causation.  Observational studies (non-uniform or biased data), on the other hand, are much less convincing because (conditional) exchangeability may be violated.

%The SUTVA, consistency, positivity, and conditional exchangeability 
 
% most models in the field of RS \col{deal with the data of observational study rather than RCT experiments}, which makes Assumptions 1-5 might not be guaranteed. 
%When one of these recoverability conditions does not hold, there are other possible approaches to making causal inference from observational data,   
% the analogy between observational study and conditionally randomized experiment breaksdown. In that situation, 

These assumptions can be divided into associational assumptions and causal assumptions~\cite{Pearl-2009}. The former (e.g. model specification of propensity score), is testable in principle. In contrast, the latter, such as Assumptions 1--5, cannot be directly verified from data, unless one resorts to experimental control. In addition, Assumptions 1--5 might not be guaranteed in observational studies, which require different sets of assumptions, such as introducing an instrumental variable or using the front-door criterion to achieve recoverability. In practice, whether the causal assumptions hold need to be discussed by expert's knowledge (e.g. drawing causal graphs) for each specific problem. 

% ==================================================================
% \section{New perspective of biases in recommender system}  % recommender system

% The main impediment of causal inference is the existence of various biases.  

% are the most common set of recoverability conditions   in causal inference.}
% Violating Assumptions 1-5 may result in various biases. Although these biased have been described in some literature in RS community~\cite{Chen-etal2020},  the formal definitions are still not clear.   This greatly limits the development of new debiasing algorithms.   
% Each step \pw{in the proposed causal analysis framework} relies on some assumptions and violating these assumptions would induce different types of biases.
% In RS, various biases are defined in a descriptive way~\cite{Chen-etal2020}, 
% In this section, we define the biases formally from a new perspective of violating what assumptions adopted in the causal analysis; see Table \ref{tab1}.
 % and relate the biases in RS.  
% However, the definitions of the biases in RS are descriptive,  formal definitions of the biases are still not clear,

\subsection{Selection Bias and Confounding Bias in RS} 
%Cause-and-effect relations are one of the most valuable types of knowledge sought after throughout the data-driven sciences since they translate into stable and generalizable explanations as well as efficient and robust decision-making capabilities.  Inferring these relations from data, however, is a challenging
%task. \pw{} 
In causal inference, selection bias and confounding bias are two of the  most common barriers to achieving causal estimates~\cite{Correa-etal2019},  which can be formally defined as violations of exchangeability and random sampling assumptions respectively. In RS, it is worth emphasizing that the research studies should first distinguish the two biases.
\begin{definition} 
	 [Selection bias] Selection bias means that the 
	  sample distribution is different from that of target population~\cite{Hernan-Robins2020}, i.e., 
		\[     \P( x, t, y ) \neq  \P_{\cO}(x, t, y).     \]  		
\end{definition}

\begin{definition}
	[Confounding bias] 
Confounding bias refers to the association ($T$ and $Y$) created due to the presence of factors affecting both the treatment and the outcome~\cite{Correa-etal2019}, i.e., $\exists t \in \cT$, $Y(t)\not\perp T$. Usually it will lead to
   \[  \bE[Y(t)] \neq \bE[Y(t)| T= t].     \]
\end{definition}

% It is noteworthy that the definition of selection bias in~\cite{Chen-etal2020} is  different from that in this article. We use ``user selection bias'' and ``selection bias'' to distinguish them.   
% significantly 
% As discussed in Section 4.1, 

Selection bias abounds in RS. For example, the system aims to recommend items that the user may like by filtering out items with low predicted ratings, and this kind of selection is previously called \textbf{model selection bias}~\cite{CIKM-YuanHYZCDL19}; users tend to rate recommended items that they like and rarely rate recommended items that they dislike, which is the \textbf{user self-selection bias}~\cite{saito2020asymmetric}. 
In such cases, the data-gathering process will reflect a distortion in the sample's proportions, since the data is no longer a faithful representation of the target population, and biased estimates will be produced regardless of the size of samples collected. 
  Interestingly, 
  %according to the descriptive definition of biases,  user selection bias, and 
  the exposure bias will also lead to   % the popularity bias, 
 $\P(x, t, y) \neq \P_{\cO}(x, t, y)$, hence belonging to the selection bias. Similar insight is founded in~\cite{Chen-etal2021}.
 % which also takes position bias as a kind of selection bias.   
 % We will discuss the problem of selection bias and confounding bias {in scenarios 1 and 2 of Section 5}.  

% Confounding bias is often viewed as the main shortcoming of observational studies. In observational studies, the treatment and outcome may be determined by many factors, then the effects of those factors become entangled (or confounded) with the effect of treatment, leading to the existence of significant differences between the distributions of different treatment groups. 
% Confounding bias cannot be eliminated as the sample size increases~\cite{Hernan-Robins2020}.  
  %Thus, the old adage association is not causation holds even if the study population is arbitrarily large 

% (even when the treatment is controlled).  
	 % Another common example is the missing data problem. In movie recommendation,  

%Passively waiting for volunteers signing up for a new experiment could not solve the problem of selection bias, even if the treatments are randomized .
Confounding bias differs fundamentally from the selection bias. Selection bias comes from the systematic bias during the collection of units into the sample~\cite{Bareinboim-etal2014}.  
A well-designed sampling procedure can reduce selection bias, such as recommending items to users randomly to obtain unbiased data, but the cost is extraordinarily expensive.
  In contrast, confounding bias stems from the systematic bias inherently determined by the causal mechanism (relations) among features, treatment, and outcome, irrespective of the data collection process. Randomization of treatment assignment can eliminate the effect of (unmeasured) confounding bias, but cannot remove the influence of selection bias~\cite{Correa-etal2019}.   
\section{Applying the Proposed Causal Analysis Framework to Recommendation Tasks}
In this section, we illustrate how to apply the proposed causal analysis framework to the classic recommendation tasks. Throughout, denote with $u \in \cU$ the users and with $i \in \cI$ the items, and denote $\cD = \cU \times \cI$ as the set of all user-item pairs. Let $Y_{u,i}$ be the outcome of interest for user-item pair $(u,i)$, $T_{u,i}$ be the treatment, and $X_{u,i}$ be the corresponding feature embedding of user $u$ and item $i$. For general treatment, define 
    \begin{equation} \label{mu}
     \mu_t(x)  = \bE[ Y(t) \mid X = x ], ~ t\in \cT,    
     \end{equation}
and for binary treatment, define 
    \begin{equation}
        \tau(x) = \bE[ Y(1) - Y(0) \mid X = x ],
    \end{equation}
which are the common causal estimands in RS. Based on the proposed causal analysis framework, it can be divided into the following scenarios according to different causal estimands and research perspectives, claiming that most of the recommendation tasks can be included into the following scenarios.

\subsection{Missing Not at Random (MNAR)}
This scenario focuses on the problem of missing outcome data. Consider the case of movie rating websites~\cite{Wang-Liang-Charlin-Blei2019,Wang-Liang-Charlin-Blei2020}.  Applying the proposed causal analysis framework, a unit is a user-item pair, the feature $X_{u,i}$ is the attributes of user $u$ and movie $i$, the outcome $Y_{u,i}$ is the true rating of user $u$ for movie $i$. However, the outcome suffers a problem of missing, and selection bias is induced due to the fact that the users incline to rate the movies they like. Usually, the missing mechanism is MNAR.  Let $O_{u,i}$ be the observing indicator of $Y_{u,i}$.  We consider the observing indicator as the treatment, then  $Y_{u,i}(1)$ denotes the true rating of user $u$ for movie $i$ if $O_{u,i} = 1$. The target population consists of all user-item pairs. The goal of the MNAR debiasing task is to estimate $Y_{u,i}(1)$ using feature $X_{u,i}$, i.e., the causal estimand of interest is $\mu_1(x)$.  
% The goal of the MNAR debiasing task is to \col{estimate the average loss of the prediction model on the missing completely at random (MCAR) data}. 

% The inverse propensity score (IPS) approach \cite{Schnabel-etal2016} aims to recover the distribution of all events by  weighting the clicked events with $1/p_{u,i}$

% 应该使用过去时吧.
Many studies have tried to give an accurate estimate of the $\mu_1(x)$ from the perspective of causality. \cite{Schnabel-Swaminathan2016} use the inverse propensity score (IPS) method to recover the distribution of the target population by weighting the non-missing units with propensity score, and further introduce the self-normalized IPS (SNIPS) estimator to reduce the large variance problem caused by extremely small estimated propensity scores. \cite{Wang-Zhang-Sun-Qi2019} propose the doubly robust (DR) method and the joint learning optimization technique. %  of imputation model and prediction model
Based on the DR estimator, \cite{Guo-etal2021} propose a more robust doubly robust (MRDR) estimator to further control the variance while retaining its double robustness. In addition, \cite{wang2020information} propose the counterfactual variational information bottleneck (CVIB) approach, and the core idea is to separate the task-aware mutual information term into factual and counterfactual parts and balance them. \cite{liu2021mitigating} propose debiased information bottleneck (DIB) based on the debiased representation from the causal diagrams and information theory.  \cite{Wu-Li-etal2022} propose a doubly robust collaborative targeted learning that makes the recommendation model more accurate and robust.

% by leveraging the targeted maximum likelihood estimation  technique.

% here we focus on the causal effects of binary treatment, that is, the difference in potential outcomes between the treatment group and control group. 
\subsection{Binary Treatment (e.g. CTR Predication)}
This scenario discusses the case of binary treatment. Different from the MNAR scenario, this scenario has no missing outcome.   
 An typical example is advertising recommendation~\cite{Gopalan-Hofman-Blei2018,Liang-Charlin-Blei2020}, where   
   a unit is a user-item pair, the target population consists of all user-item pairs, and the outcome $Y_{u, i}$ is the indicator of a click event, 
   i.e., $Y_{u,i} = 1$ if user $u$ clicks item $i$, $Y_{u,i} = 0$ otherwise. The treatment $T_{u,i} = 1$ if item $i$ is exposed to user $u$, $T_{u,i} = 0$ otherwise, the potential outcomes $Y_{u,i}(1)$ and $Y_{u,i}(0)$ denote the indicator of click event if the item is/isn't exposed to the user $u$. The estimand of interest is $\mu_1(x)$ denoting the click-through rate (CTR),  or $\tau(x)$ denoting the uplift of CTR.
   
% Besides, it is worth noting that 
The uplift modeling in RS~\cite{Sato-Singh2019} are closely related to the binary treatment scenario, which aims to predict the change of feedback value caused by the increment of the treatment, and there are many studies that have successfully applied causal inference techniques to estimate the causal effect in uplift models accurately~\cite{gutierrez2017causal,athey2015machine,hitsch2018heterogeneous}.

\subsection{Multi-valued Treatment (e.g. Position Bias)}
Consider the scenario of multi-valued treatment, which corresponds to the contextual bandit in RS with finite action space \cite{li2010contextual}. We also consider the task of CTR prediction and assume it suffers from the problem of position bias~\cite{yuan2020unbiased}. 
If the exposed locations are available, assuming that there are $K$ different positions, then this is a problem of multi-valued treatment. Specifically, the treatment $T_{u,i}$ has $K$ levels, where  
$T_{u,i}= j$ means that the item $i$ is exposed to user $u$ at $j$-th position. Correspondingly, there are $K$ potential outcomes for each unit. 
The causal estimand can be $\mu_j(x)$ or $\tau_{j,k}(x) = \bE[ Y_{u, i}(j)-Y_{u, i}(k) \mid X_{u, i} = x]$. The former represents the CTR at position $j$, while the latter indicates the change in CTR from exposure to position $k$ to position $j$.
% we can treat the position as a treatment. 
% we can treat  each item as a unit, the treatment is the location of the item, and the potential outcome is the  click label when the item is displayed in a specific location.

\subsection{Continuous Treatment (e.g. Cash Reward)}
 % Recommending ads with rewards. 
This section further extends the scenarios in Sections 5.2 and 5.3 to continuous treatment, which corresponds to the contextual bandit in RS with infinite action space \cite{li2010contextual}. In fact, in order to increase CTR or post-click conversion rate (CVR), many RS (like TikTok and Kuaishou) carry cash rewards when users click on an advertisement, in which the reward is usually a continuous variable. In this situation, we treat the user-item pairs as units representing the scenario of user $u$ clicking the ads of item $i$, reward $T_{u, i}$ as the continuous exposure treatment, and the corresponding profit $Y_{u, i}$ as outcomes. 
If the goal of RS is to estimate the cumulative net income, the estimand of interest is $\mu_t(x) - t=\bE[Y_{u, i}(t) - t \mid X_{u, i}=x]$, where $t$ is the treatment value, representing the cash rewards for $u$ clicking the ads of item $i$.

\subsection{Compliance (e.g. Exposure-Click-Conversion)}
The compliance scenario involves two variables $C$ and $Y$ measured after the treatment $T$, and the causal relationship among them is $T \to C \to Y$, i.e., $T$ affects $Y$ only through $C$. Consider an example of online advertising. The units are the user-item pairs, the target population is all the user-item pairs, and the feature $X_{u,i}$ is the embedding of user $u$ and item $i$. $T_{u,i}$, $C_{u,i}$, and $Y_{u,i}$ are indicators of the user exposed on the item advertising, click on the item and the conversion on the item.  
If we treat $T$ as  treatment, then $\bE[ C(1) \mid X = x]$ denotes the CTR. If we treat $C$ as treatment, then $\bE[ Y(1) \mid X = x]$ denotes the CVR. However, if we want to detect the effect of $T$ on $Y$, then the estimand is more complicated.   In such a case, we regard $T$ as the treatment and let $C(0)$ and $C(1)$ be the potential click behaviors if $T=1$ and $T=0$, respectively. The definition of potential conversion depends on both $T$ and $C$. Let $Y(t, c)$ be the potential conversion if $T = t$ and $C(t)= c$.  Then the estimand       
    \begin{equation}
     \tilde \tau(x) =  \bE[ Y(1, C(1) ) - Y(0, C(0)) \mid X  = x ] 
    \end{equation}    
measures the causal effect  of $T$ on $Y$.

The idea of compliance is widely used in causal inference, but it is rarely discussed in RS. \cite{ADKDD-tgkkhz21} considers the compliance framework from the perspective of advertising demanders. Since click advertising requires payment, we prefer to push advertising to user-item pairs with causal effects rather than free rider (referring to user-item pairs that will always be converted regardless of whether advertising is recommended or not), and this phenomenon is very common in popular products.

\subsection{Recommendation Policy Evaluation and Learning}
 This scenario treats the recommendation problem as a policy learning problem, and no longer pays attention to the estimands $\mu_t(x)$ and $\tau(x)$. 
For reinforcement learning in RS, it is always focusing on the evaluation or optimization of recommendation strategies \cite{afsar2021reinforcement,munemasa2018deep}. The counterfactual framework can be further used to deal with the delayed feedback \cite{zhang2021counterfactual}, which acts as a common research direction in RS. From the causal perspective, suppose that there are a total of $I$ items and $U$ users. A unit is a user, the feature $X_u$ is the attribute of user $u$, the treatment $T_u$ has $I$ levels, denoted as $\cT = \{1, 2, \cdots, I \}$, where $T_u = i$ means that item $i$ is exposed to user $u$.  
The reward caused by user $u$ exposed to the item $i$ as the potential outcome is denoted by $Y_u(i)$. The target population is all the users. 
The observed data consist of $U$ observations
of features $X_u$, treatment $T_u$, and reward $Y_u = Y_u(T_u)$. And the target quantity is the optimal policy defined by  
  $$   \pi_0^* = \arg \max_{\pi \in \Pi_0} V(\pi), $$
where $\Pi_{0}$ is a policy class, $V(\pi)$ is the policy value, refers to the expectation of the reward under the policy $\pi$, i.e., 
    \[  V(\pi) =\bE\left[ \sum_{t \in \mathcal{T} }\pi(t | X) Y(t) \right] = \bE\left[ \sum_{t \in \mathcal{T} }\pi(t | X) \mu_t(X) \right],    \]
where $\mu_t(x)$ is defined in (\ref{mu}).

% {causal estimands in Sections 5.1 to 5.6 under \col{Assumptions 1-5} (\pw{section 5.6 is different.}), that is, there is no selection bias and the main challenge of estimating $\tau(x)$ is the confounding bias \pw{Section 5.1 is different}.}
 %In addition, there are a large number of other methods. Here we introduce a few of them for illustration, as the estimating approaches are not the main focus of this article. 
\subsection{Existing Causal Debiasing Methodologies} 
The methods developed in the MNAR scenario can be applied for the estimation of $\mu_t(x)$ and $\tau(x)$ in Sections 5.2 to 5.6. Besides, this section further reviews other existing debiasing methods under Assumptions 1-5. 
Throughout, let $\mu(x) = \bE[Y|X=x]$, $\pi(x) = \P(T=1|X=x)$, $\mu_t(x) = \bE[Y(t) |X=x] = \bE[Y|X=x, T=t]$ for $t = 0, 1$.% and assumption 2-4 holds.

Many statistical methods can be used to estimate the causal effect, including S-learner \cite{Hill2011}, T-learner \cite{Hansotia-Rukstales2001}, U-learner \cite{Nie-Wager2021}, R-learner \cite{Nie-Wager2021}, X-learner \cite{Kunzel-etal2019}, IPW-learner \cite{Horvitz-Thompson1952} and DR-learner \cite{Kennedy-2020}. For example, the U-learner obtains the estimator of $\tau(x)$ by regressing $(Y-\mu(X))/(T-\pi(X))$ on $X$, and IPW-learner by regressing $T Y / \pi(X)  -  (1- T) Y  /( 1 - \pi(X) )$ on $X$. Clearly, these methods rely on the model specifications of the nuisance parameters $\pi(X)$, $\mu_t(X)$ or $\mu(X)$.  
There are many other machine learning methods that are designed to estimate $\tau(x)$ directly, such as causal tree \cite{Athey-Imbens-2015}, causal forest \cite{Wager-Athey-2018,Lechner-2019,Athey-Tibshirani-Wager-2019,Oprescu-etal-2020}, causal BART \cite{Hahn-Murray-Carvalho2020}, causal boosting and causal MARS \cite{powers2018some}, balancing counterfactual regression \cite{johansson2016learning},  Generative Adversarial Nets \cite{yoon2018ganite}, Causal Effect Variational Autoencoder \cite{louizos2017causal}, and local similarity preserved SITE \cite{yao2018representation}.

% ====================================================
\section{Open Research Directions}
Recently, more and more researchers in RS are trying to apply causal inference methods to handle RS tasks such as CTR/CVR prediction, delayed feedback, etc., nevertheless, there are still many challenges and opportunities. By matching the existing research with the causal analysis framework discussed above, we have identified the following open research directions, which are rarely formalized in the potential outcome framework for RS to conduct research.

 \paragraph{\emph{Data Fusion.}} A typical scenario involves a combination of a large biased (observational or non-uniform) dataset and a small unbiased (experimental or uniform) dataset. The biased data and unbiased data have complementary characteristics. The biased data is inevitable to suffer from the problem of hidden/unmeasured confounders, which will distort the causal conclusions even the sample size is infinity \cite{Kallus-Puli2018}. 
  In comparison, collected through a carefully designed experiment, the unbiased data has no (hidden) confounding bias, and it provides the gold standard for evaluating the debiasing approaches. In summary, data fusion is a promising strategy to improve the quality of RS.
  
 \paragraph{\emph{Sequential Recommendation.}} By modeling the user behavior sequence, such as the sequence of purchasing items, RS can learn the change of user interest and predict the user's next behavior. From the perspective of causality, it can be considered that the assignment mechanism and potential outcomes of RS are changing with the time series, and the goal is to dynamically capture the changes of users' interests, so as to achieve more accurate recommendations.
   
   \paragraph{\emph{Fairness in RS.}} Many literature define group fairness and individual fairness through counterfactual causality \cite{kusner2017counterfactual,nabi2018fair,chiappa2019path}. However, how to formalize the fairness in RS with causal framework is still vague, especially when the user has a social network, which will violate SUTVA(b), and bring greater challenges to the well-definedness of causal fairness. In addition, there is still a lot of research space on how to modify the traditional causal recommendation model to achieve the balance of accuracy and fairness.
   % \item Causal Analysis with Unobserved Confounder:
  
   \paragraph{\emph{Interference.}} Even though almost all articles on causality inspired recommendation acquiesce in the SUTVA assumption, as discussed in Section 3, the SUTVA assumption will be violated in many cases, resulting in biased estimation. Another form of interference is between potential outcomes of different units. For example, a user's purchase behavior will affect the purchase behavior of other users in his social network, which is often encountered in social recommendation. 

\section{Conclusion}
Causality offers new opportunities for robust and outstanding performance of algorithms for debiasing and prediction tasks in RS. This article reviews related research by providing a unified causal analysis framework for RS, revealing and discussing in detail the validity of the always neglected but equally important causal assumptions. New interpretations of various biases in RS are provided from the perspective of violating causal assumptions. The proposed causal RS analysis framework is applied to rigorously formulate a large number of RS tasks, such as the non-compliance problem, interference bias, and policy learning, which have been intensively studied in causal inference literature. 
% strictly

The paper concludes with an overview of causal estimation methods that will hopefully provide new research opportunities in the field of causal recommendation including but not limited to debiasing and prediction tasks. In addition, it is expected to develop new methods with weakening or substituting the common assumptions in RS studies. 

\newpage

\appendix

\bibliographystyle{named}
\bibliography{refs}

\end{document}